\documentclass[pra,aps,superscriptaddress,amssymb,amsmath,reprint,noeprint,floatfix]{revtex4-2}
\usepackage{natbib}
\usepackage{braket}
\usepackage{bm}
\usepackage{graphicx}
\usepackage{hyperref}
\usepackage{xcolor}
\hypersetup{bookmarksnumbered}

\newcommand{\EqnOne}{
    \begin{equation}
        \begin{aligned}
        \mathbf{E}_{\rm s}={\rm E}_0\sum_{l=1}^\infty\sum_{m=-l}^li^l\left[\pi\left(2l+1\right)\right]^\frac{1}{2}\Bigl\{\frac{1}{k}a_{\rm E}\left(l,m\right)\nabla\times\\
        \left[h_l^{\left(1\right)}(kr)\mathbf{X}_{lm}(\theta,\phi)\right]+a_{\rm M}(l,m)h_l^{\left(1\right)}(kr)\mathbf{X}_{lm}(\theta,\phi)\Bigr\},
        \end{aligned}
        \label{eqn:1}
    \end{equation}
}

\newcommand{\EqnTwo}{
    \begin{equation}
        \begin{gathered}
        a_{\mathrm{E}}(l, m)=\frac{(-i)^{l-1} k^{2} \eta O_{l m}}{\mathrm{E}_{0}[\pi(2 l+1)]^{\frac{1}{2}}} \int \exp (-i m \varphi)\Bigl\{\bigl[\psi_{l}(k r) \\
        +\psi_{l}^{\prime \prime}(k r)\bigr] P_{l}^{m}(\cos \theta) \hat{r} \cdot \mathbf{J}_{\mathrm{s}}(\mathbf{r})+ \\
        \frac{\psi_{l}^{\prime}(k r)}{k r}\left[\tau_{l m}(\theta) \hat{\theta} \cdot \mathbf{J}_{\mathrm{s}}(\mathbf{r})-i \pi_{l m}(\theta) \hat{\varphi} \cdot \mathbf{J}_{\mathrm{s}}(\mathbf{r})\right]\Bigr\} \mathrm{d}^{3} r,
        \end{gathered}
        \label{eqn:2}
    \end{equation}
}

\newcommand{\EqnThree}{
    \begin{equation}
        \begin{gathered}
        a_{\mathrm{M}}(l, m)=\frac{(-i)^{l+1} k^{2} \eta O_{l m}}{\mathrm{E}_{0}[\pi(2 l+1)]^{\frac{1}{2}}} \int \exp (-i m \varphi) j_{l}(k r) \\
        \cdot\left[i \pi_{l m}(\theta) \hat{\theta} \cdot \mathbf{J}_{\mathrm{s}}(\mathbf{r})+\tau_{l m}(\theta) \hat{\varphi} \cdot \mathbf{J}_{\mathrm{s}}(\mathbf{r})\right] \mathrm{d}^{3} r,
        \end{gathered}
        \label{eqn:3}
    \end{equation}
}

\newcommand{\EqnFour}{
    \begin{equation}
        \begin{aligned}
            C_{\mathrm{sca}}=\frac{\pi}{k^{2}} \sum_{l=1}^{\infty} \sum_{m=-l}^{l}(2 l+1)\left(\left|a_{\mathrm{E}}\right|^{2}+\left|a_{\mathrm{M}}\right|^{2}\right).
        \end{aligned}
        \label{eqn:4}
    \end{equation}
}

\newcommand{\EqnFive}{
    \begin{equation}
           \chi=\frac{\varepsilon_{0}}{2} \mathcal{E} \cdot \nabla \times \mathcal{E}+\frac{1}{2 \mu_{0}} \mathcal{B} \cdot \nabla \times \mathcal{B}=-\frac{\varepsilon_{0} \omega}{2}\operatorname{Im}\left({\mathbf{E}}^{*} \cdot {\mathbf{B}}\right),
        \label{eqn:5}
    \end{equation}
}

\newcommand{\EqnSix}{
        \begin{equation}
        \begin{aligned}
        g=\frac{2\left(A^\mathrm{L}-A^\mathrm{R}\right)}{A^\mathrm{L}+A^\mathrm{R}},
        \end{aligned}
        \label{eqn:6}
    \end{equation}
}

\newcommand{\EqnSeven}{
    \begin{equation}
        \begin{aligned}
        A=\frac{\omega}{2} \mathrm{Im}\left(\mathbf{E}^{*} \cdot \mathbf{p}+\mathbf{B}^{*} \cdot \mathbf{m}\right),
        \end{aligned}
        \label{eqn:7}
    \end{equation}
}

\newcommand{\EqnEight}{
    \begin{equation}
        \begin{aligned}
        \mathbf{p}&=\overleftrightarrow{\alpha}_{\mathrm{e}} \mathbf{E}+\mathrm{i} \overleftrightarrow{\kappa} \mathbf{B},\\ \mathbf{m}&=\overleftrightarrow{\alpha}_{\mathrm{m}} \mathbf{B}-\mathrm{i} \overleftrightarrow{\kappa}^{\mathrm{T}} \mathbf{E},
        \end{aligned}
        \label{eqn:8}
    \end{equation}
}

\newcommand{\EqnNine}{
    \begin{equation}
        \begin{aligned}
        A^{\mathrm{L}}-A^{\mathrm{R}}
        =&\frac{\omega}{2} \mathrm{Im}[\mathbf{E}^{*}\cdot\left(\overleftrightarrow{\alpha}_{\mathrm{e}}^{\mathrm{L}}-\overleftrightarrow{\alpha}_{\mathrm{e}}^{\mathrm{R}}\right)\cdot\mathbf{E}\\
        &+\mathbf{B}^{*}\cdot\left(\overleftrightarrow{\alpha}_{\mathrm{m}}^{\mathrm{L}}-\overleftrightarrow{\alpha}_{\mathrm{m}}^{\mathrm{R}}\right)\cdot\mathbf{B}]\\
        &+\mathrm{i} 2 \mathbf{E}^{*}\cdot\left(\overleftrightarrow{\kappa}^{\mathrm{L}}-\overleftrightarrow{\kappa}^{\mathrm{R}}\right)\cdot\mathbf{B}]].
          \end{aligned}
        \label{eqn:9}
    \end{equation}
}

\newcommand{\EqnTen}{
    \begin{equation}
   A^{\mathrm{L}}+A^{\mathrm{R}} = \frac{\omega}{2} \operatorname{Im}\left[\mathbf{E}^{*} \cdot\left(\overleftrightarrow{\alpha}_{\mathrm{e}}^{\mathrm{L}}+\overleftrightarrow{\alpha}_{\mathrm{e}}^{\mathrm{R}}\right) \cdot \mathbf{E}\right],
    \label{eqn:10}
    \end{equation}
}

\newcommand{\EqnEleven}{
    \begin{equation}
A^{\mathrm{L}}+A^{\mathrm{R}} = \frac{4 \omega}{\varepsilon_{0}} \alpha_{\mathrm{e}}^{\prime \prime} U_{\mathrm{e}},
    \label{eqn:11}
    \end{equation}
}

\newcommand{\EqnTwelve}{
    \begin{equation}
g = \frac{\varepsilon_{0} \operatorname{Im}\left[\mathbf{E}^{*} \cdot\left(\overleftrightarrow{\alpha}_{\mathrm{e}}^{\mathrm{L}}-\overleftrightarrow{\alpha}_{\mathrm{e}}^{\mathrm{R}}\right) \cdot \mathbf{E}\right]}{4 \alpha_{\mathrm{e}}^{\prime \prime} U_{\mathrm{e}}}-\sum_{i} \frac{2 \kappa_{i i}^{\prime \prime} \chi_{i}}{\omega \alpha_{\mathrm{e}}^{\prime \prime} U_{\mathrm{e}}}.
    \label{eqn:12}
    \end{equation}
}
\newcommand{\FigOne}{
    \begin{figure}[t!]
        \centering
        \includegraphics[width=\linewidth]{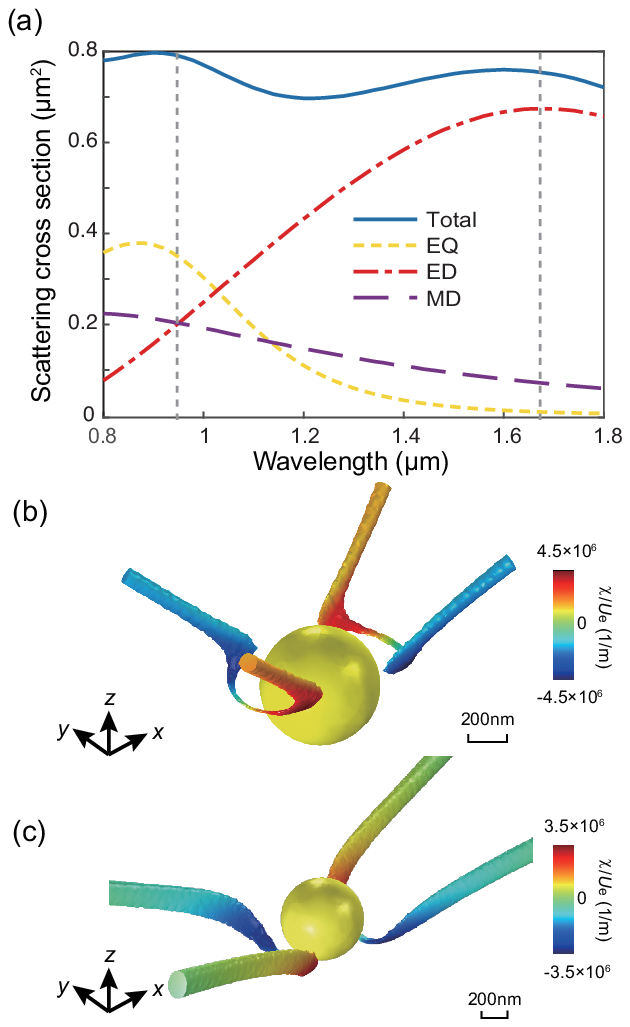}
        \caption{(a) Scattering cross section of the gold sphere illuminated by a \textit{x}-polarized plane wave propagating in \textit{z} direction. (b) Type-I C line of the gold sphere at $\lambda=950$ nm. (c) Type-II C line of the gold sphere at $\lambda=1680$ nm. The color of the C lines denotes the normalized optical chirality $\chi/U_\text{e}$.}
        \label{fig:1}
    \end{figure}
}

\newcommand{\FigTwo}{
    \begin{figure}[t!]
        \centering
        \includegraphics[width=\linewidth]{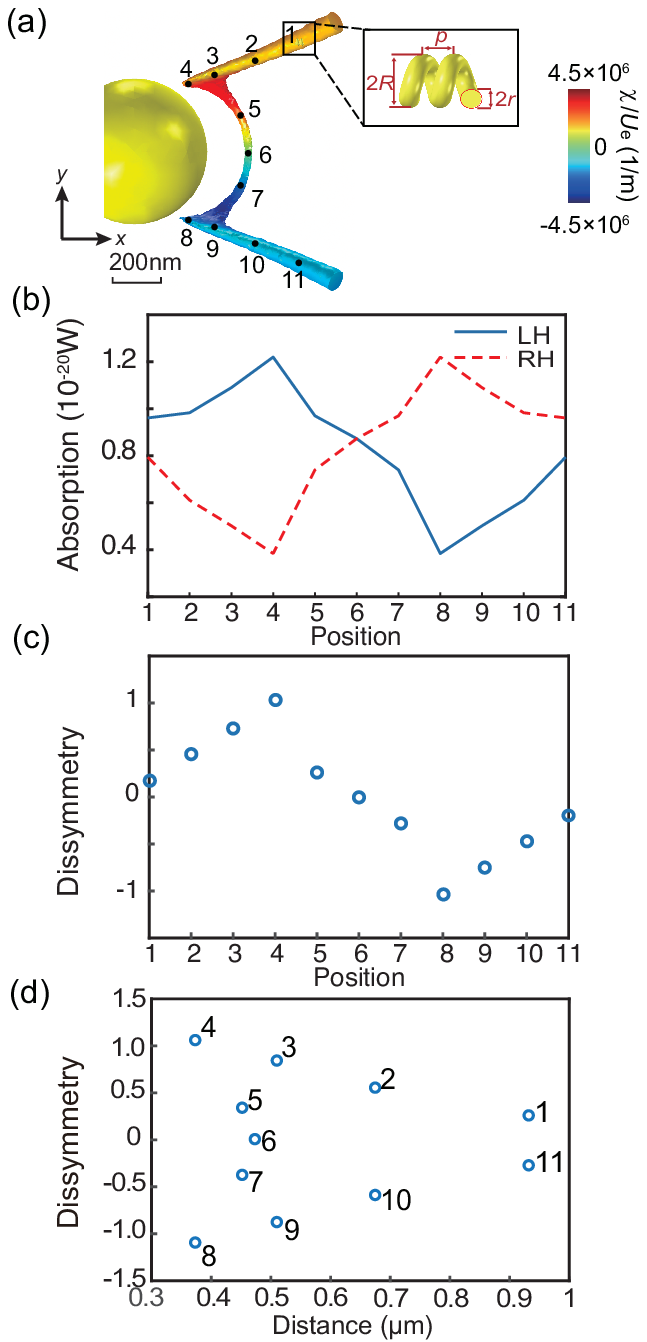}
        \caption{(a) Schematic for chiral discrimination by the Type-I C line. A two-pitch gold helix serves as the chiral particle and is located at the positions labeled as 1-11. Its inner radius $r = 5$ nm, outer radius $R = 10$ nm, pitch $p = 15$ nm. (b) The absorption of the LH and RH helices as a function of the position. (c) The dissymmetry factor as a function of the helix’s position. (d) The dissymmetry factor as a function of distance from the gold nanoparticle.}
        \label{fig:2}
    \end{figure}
}

\newcommand{\FigThree}{
    \begin{figure}[t!]
        \centering
        \includegraphics[width=\linewidth]{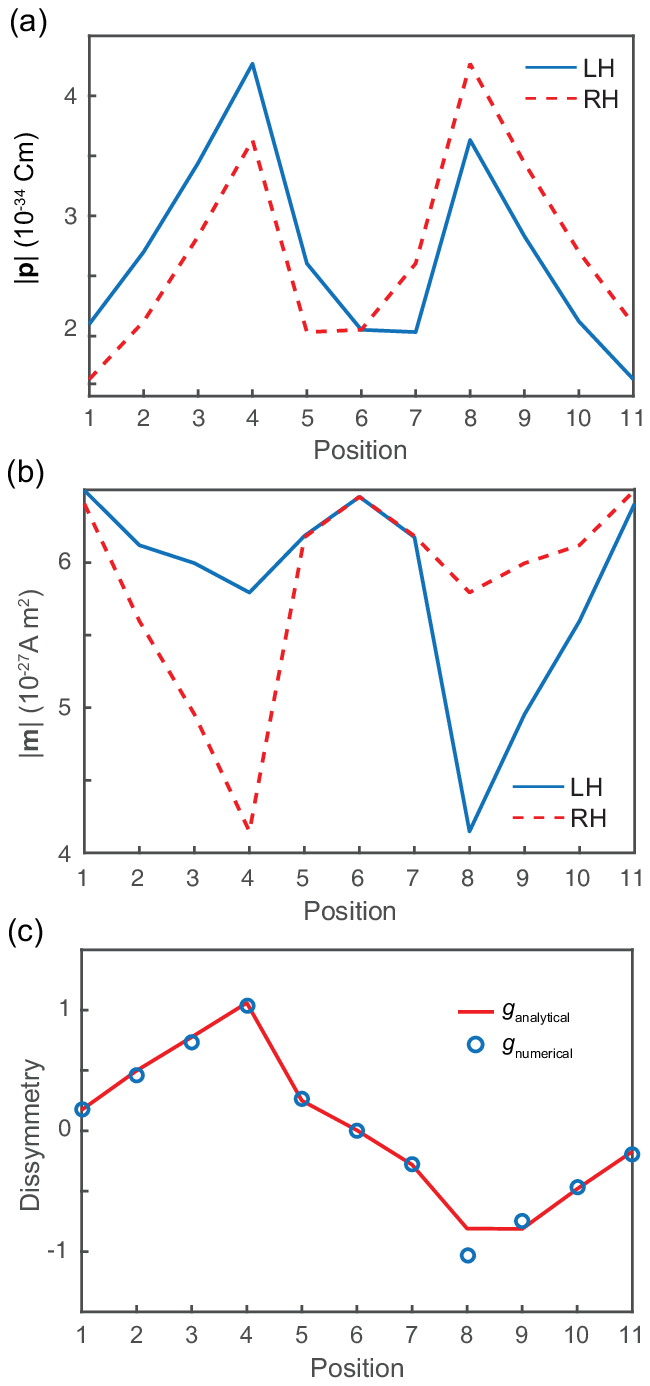}
        \caption{Amplitude of (a) the electric dipole moment \textbf{p} and (b) the magnetic dipole moment \textbf{m} as a function of the position for the Type-I C line. (c) Comparison between the analytical and numerical  results of the dissymmetry factor.}
        \label{fig:3}
    \end{figure}
}

\newcommand{\FigFour}{
    \begin{figure}[t!]
        \centering
        \includegraphics[width=\linewidth]{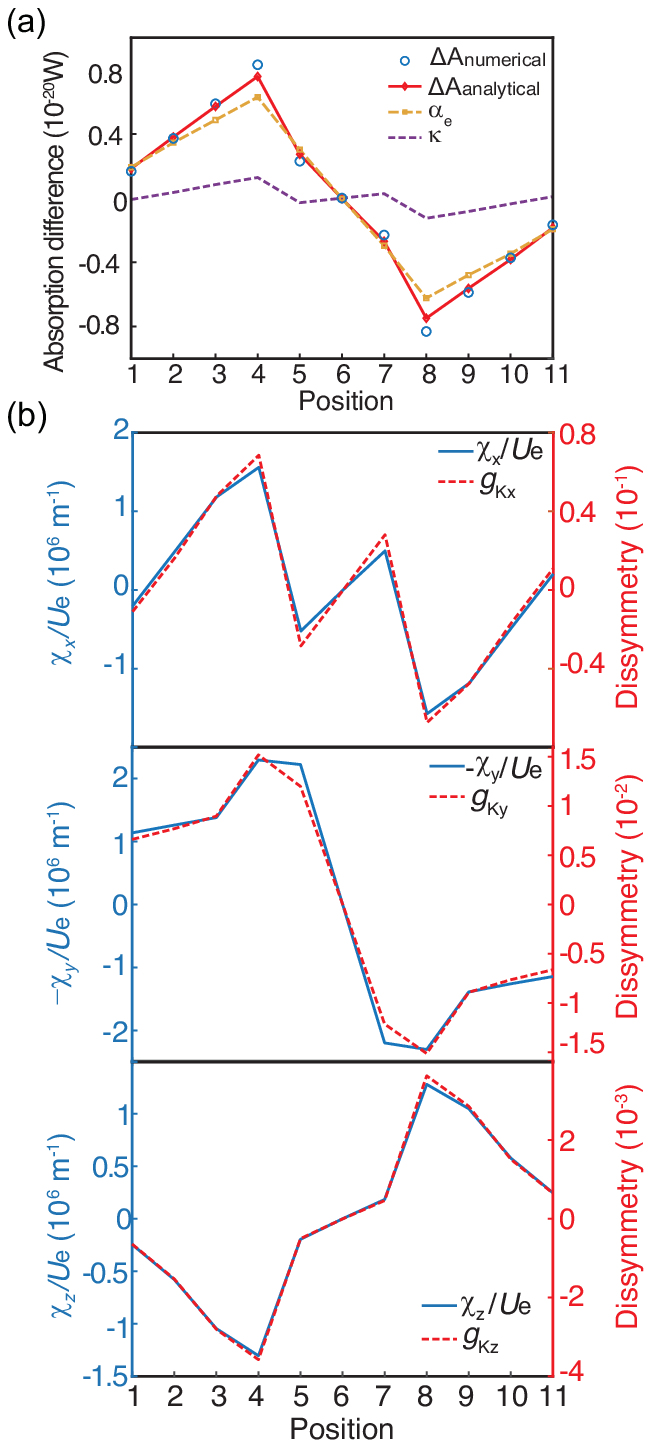}
        \caption{(a) The numerical and analytical results of absorption difference and the contributions of electric and magnetoelectric polarizabilities in the case of Type-I C line. (b) The dissymmetry factor and normalized optical chirality contributed by each Cartesian component of the field.}
        \label{fig:4}
    \end{figure}
}
\newcommand{\FigFive}{
    \begin{figure}[t!]
        \centering
        \includegraphics[width=\linewidth]{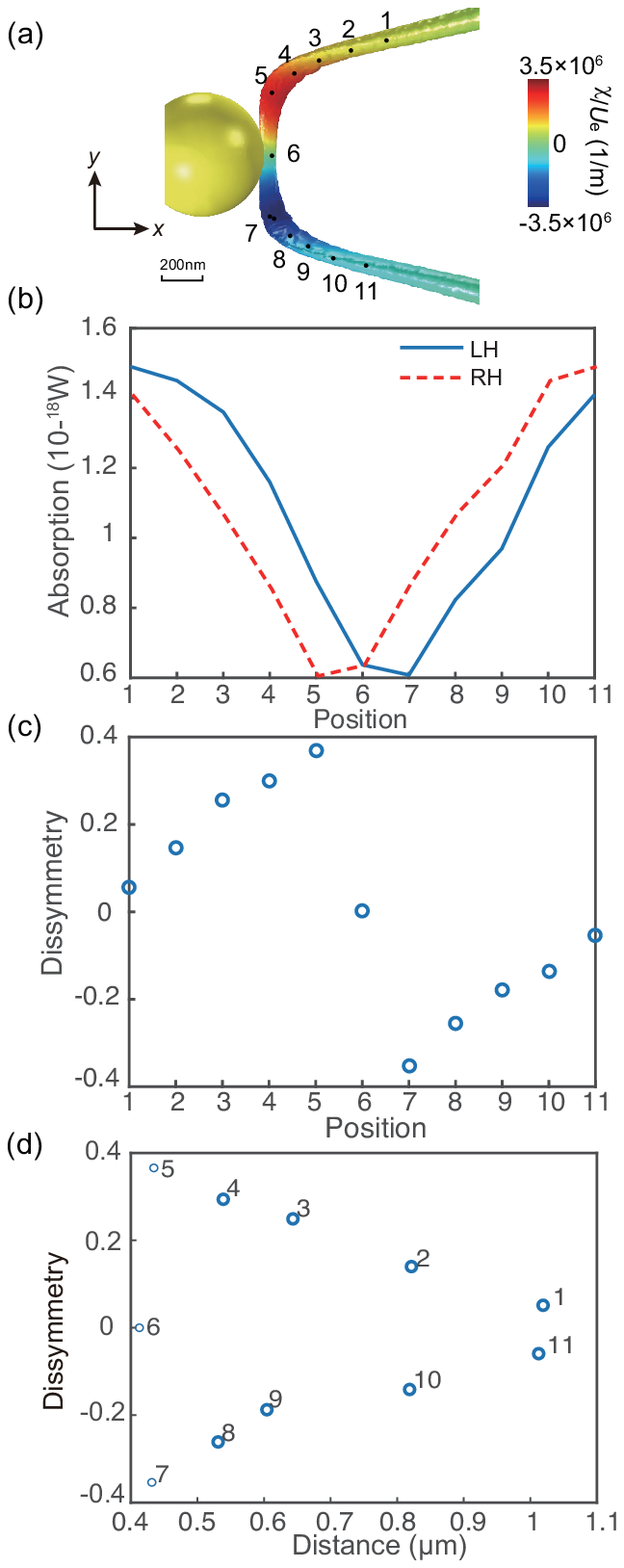}
        \caption{(a) Schematic for chiral discrimination by the Type-II C line. (b) The absorption of the LH and RH helices as a function of the position. (c) The dissymmetry factor as a function of the helix’s position. (d) The dissymmetry factor as a function of distance from the gold nanoparticle.}
        \label{fig:5}
    \end{figure}
}

\newcommand{\FigSix}{
    \begin{figure}[t!]
        \centering
        \includegraphics[width=\linewidth]{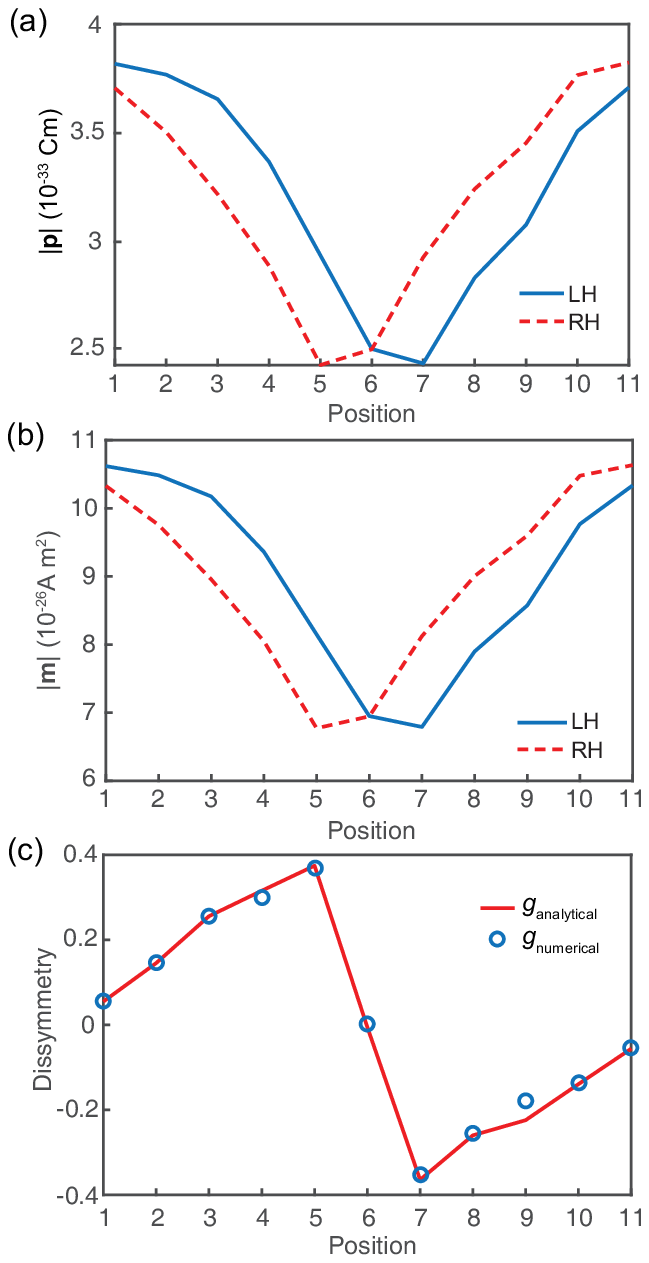}
        \caption{Amplitude of (a) the electric dipole moment \textbf{p} and (b) the magnetic dipole moment \textbf{m} as a function of the position for the Type-II C line. (c) Comparison between the analytical and numerical results of the dissymmetry factor.}
        \label{fig:6}
    \end{figure}
}

\newcommand{\FigSeven}{
    \begin{figure}[t!]
        \centering
        \includegraphics[width=\linewidth]{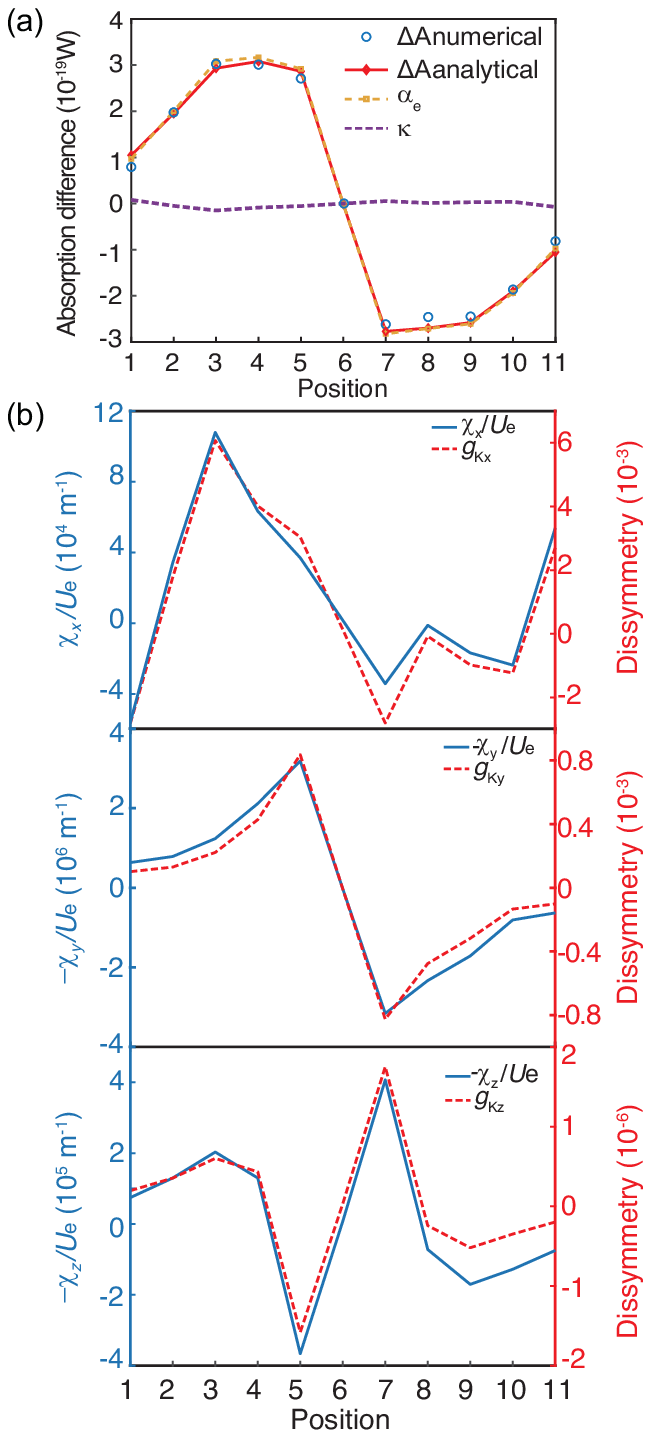}
        \caption{(a) The numerical and analytical results of absorption difference and the contributions of electric and magnetoelectric polarizabilities in the case of Type-II C line. (b) The dissymmetry factor and normalized optical chirality contributed by each Cartesian component of the field.}
        \label{fig:7}
    \end{figure}
}

\begin{document}

\title{Chiral discrimination by polarization singularities of a metal sphere}
\date{\today}

\author{Shiqi Jia}
\affiliation{Department of Physics, City University of Hong Kong, Hong Kong, China}
\author{Jie Peng}
\affiliation{Department of Physics, City University of Hong Kong, Hong Kong, China}
\author{Yuqiong Cheng}
\affiliation{Department of Physics, City University of Hong Kong, Hong Kong, China}
\author{Shubo Wang}\email{shubwang@cityu.edu.hk}
\affiliation{Department of Physics, City University of Hong Kong, Hong Kong, China}
\affiliation{City University of Hong Kong Shenzhen Research Institute, Shenzhen, Guangdong 518057, China}

\begin{abstract}
 The practical applications of chiral discrimination are usually limited by the weak chiral response of enantiomers and the high complexity of detection methods. Here, we propose to use the C lines (i.e., lines of polarization singularities) emerged in light scattering by a metal sphere to detect small chiral particles. Using full-wave numerical simulations and analytical multipole expansions, we determined the absorption dissymmetry of deep-subwavelength helices (i.e., chiral particles) at different positions on the C lines. We find that the absorption dissymmetry can be much larger than that induced by circularly polarized plane wave excitation, and it can be attributed to the asymmetric absorption of the induced electric and magnetic dipoles. \textcolor{black}{We also show that the absorption dissymmetry strongly depends on the anisotropy of the helices}. The results  may find applications in optical manipulations, optical sensing, and chiral quantum optics. 
\end{abstract}
\maketitle

\section{\label{sec: I. Introduction}Introduction}
Chirality is a universal property of symmetry in nature that plays important roles in physics, chemistry, and biology \cite{barronMolecularLightScattering2004}. An object possesses chirality if it cannot be superimposed on its mirror image. A chiral particle and its mirror-imaged particle form a pair of enantiomers that can have dramatic different properties deriving from opposite chirality \cite{melchiorreAsymmetricAminocatalysisGold2008,wangLateralOpticalForce2014,zhangOpticalForceToroidal2015,chenMechanicalEffectPhotonic2018,panHighlyEnantioselectiveSynthesis2019,woOpticalForcesCoupled2020}. The detection/separation of different enantiomers, referred to as chiral discrimination, has long been a major interest due to important applications in various fields such as pharmaceutical industries \cite{eastgateDesignComplexDrug2017,sudaLightdrivenMolecularSwitch2019}.

One widely used chiral discrimination technique is circular dichroism (CD) spectroscopy  \cite{barronMolecularLightScattering2004}, which has been applied to detect chiral molecules \cite{freedmanAbsoluteConfigurationDetermination2003}, proteins \cite{tulliusSuperchiralSpectroscopyDetection2015}, nanostructures  \cite{vinegradCircularDichroismSingle2018} and so on. The CD is based on the differential absorption of enantiomers under left-circularly polarized (LCP) light and right-circularly polarized (RCP) light. It is typically orders of magnitude weaker than the usual absorption spectrum. This is because that CD requires the coupling between electric and magnetic dipoles \cite{powerCircularDichroismGeneral1974}, while the usual absorption only requires electric dipole and thus is the much stronger. To enhance the chiral signals and the sensitivity of chiral discrimination, various methods have been proposed, such as using the superchiral field of a standing wave  \cite{tangOpticalChiralityIts2010}, vector beams with strong longitudinal components \cite{yeEnhancingCircularDichroism2021}, chiral microwave three-wave mixing \cite{pattersonSensitiveChiralAnalysis2013}, plasmonic nanostructures \cite{govorovPlasmonInducedCircularDichroism2011,zambrana-puyaltoAngularMomentuminducedCircular2014,hoEnhancingEnantioselectiveAbsorption2017,hentschelChiralPlasmonics2017,lasaalonsoSurfaceEnhancedCircularDichroism2020,garcia-etxarriFieldMediatedChiralityInformation2020}, and metasurfaces \cite{hendryUltrasensitiveDetectionCharacterization2010,garcia-guiradoEnantiomerSelectiveMolecularSensing2018,mohammadiNanophotonicPlatformsEnhanced2018,gorkunovMetasurfacesMaximumChirality2020}. Photoexcitation circular dichroism was also shown to have orders of magnitude improvement on the sensitivity \cite{beaulieuPhotoexcitationCircularDichroism2018}. 

In this paper, we propose a new method of chiral discrimination by using the C lines emerged in light scattering by a metal sphere. The C lines correspond to the loci of polarization singularities at which light is  circularly polarized and the major axis of the polarization ellipse is not well defined \cite{nyeWaveStructureMonochromatic1987}. The C lines have been theoretically studied in various systems including paraxial waves \cite{dennisPolarizationSingularitiesParaxial2002,berryElectricMagneticPolarization2004}, Gaussian beams \cite{freundPolarizationSingularityIndices2002}, general 3D fields \cite{berryIndexFormulaeSingular2004}, and photonic structures \cite{chenExtremizeOpticalChiralities2021}. Interestingly, they can also emerge in light scattering by simple small particles, where the C lines can extend from the near-field to the far-field region \cite{pengPolarizationSingularitiesLight2021}. It has been shown that the emergence of such polarization singularties are intimately connected to the topology of photonic structures \cite{peng2022}, and the C lines can been employed to construct novel topological configurations of light fields including Mobius strip \cite{garcia-etxarriOpticalPolarizationMobius2017,peng2022} and knots \cite{grigorievFineCharacteristicsPolarization2018,larocqueReconstructingTopologyOptical2018,sugicSingularKnotBundle2018,pisantyKnottingFractionalorderKnots2019,kuznetsovThreedimensionalStructurePolarization2020}. Here, we apply the C lines emerged in light scattering to detect chiral particles. In contrast to conventional methods of chiral discrimination, our proposed method does not rely on the use of chiral structures or chiral excitations to generate chiral fields, and it only employs an achiral metal sphere excited by a linearly polarized light. By calculating the absorption dissymmetry factor of deep-subwavelength helices (i.e., chiral particles), we show that the C lines can provide much higher sensitivity than circularly polarized plane waves. \textcolor{black}{We have considered two types of C lines that are attributed to different multipoles induced in the metal sphere, and show that the C line with higher-order multipoles can give rise to a larger absorption dissymmetry factor. In addition, we show that the absorption dissymmetry strongly depends on the anisotropy of the helix.} 

The paper is organized as follows. In Sec. II, we discuss the scattering properties of the metal sphere and the induced C lines, where two types of C lines are identified based on their multipole components. In Sec. III, we show the absorption property of a chiral particle located on the Type-I C line that is contributed by high-order multipoles. As a comparison, in Sec. IV, we show the absorption property of the chiral particle located on the Type-II C line that is dominated by electric dipole. In both two cases, we compare our numerical results with the analytical results of multipole expansions. We then draw the conclusion in Sec. V. 

\FigOne

\section{\label{sec: II. MST,SourceRep}	The C lines of a gold sphere }
We consider a gold sphere under the excitation of a linearly polarized plane wave propagating in $z$ direction and polarized in $x$ direction.  \textcolor{black}{This simple achiral configuration is easy to realize experimentally, and it can generate C lines via the interference of multipoles \cite{pengPolarizationSingularitiesLight2021}.} The radius of the sphere is $r = 300$ nm. The relative permittivity of gold is described by a Drude model $\varepsilon_\text{Au}=1-\omega_\text{p}^2\text{⁄}(\omega^2+i\omega\gamma)$, where the plasmonic frequency is $\omega_\text{p}=1.28\times10^{16}$  rad/s, and the damping frequency is $\gamma=7.10\times10^{13}$  rad/s \cite{olmonOpticalDielectricFunction2012}. To characterize the scattering property of the gold spherical particle, we first conduct full-wave simulations of the system by using COMSOL Multiphysics \cite{WwwComsolCom} and calculated the scattering cross section of the sphere. The result is denoted by the solid blue line in Fig. \ref{fig:1}(a). We then decompose the scattering field of the sphere into electromagnetic multipoles as \cite{bohrenAbsorptionScatteringLight1998,johndavidjacksonClassicalElectrodynamics2012}:
\EqnOne
where $\mathrm{E_0}$ is incident electric field amplitude, $\mathbf{X}_{lm}$ and $h_l^{(1)}$ are the normalized vector spherical harmonics and the spherical Hankel functions of the first kind, respectively. The vector functions $\nabla\times[h_l^{(1)}(kr)\mathbf{X}_{lm}(\theta,\phi)]$ and $h_l^{(1)}(kr)\mathbf{X}_{lm}(\theta,\phi)$ constitute a complete basis, where each function of $\left(l, m\right)$ describes the field contributed by a unique multipole. The integer \textit{l} indicates the order of the multipole, and the integer \textit{m} denotes the projection of the angular momentum along \textit{z} axis. The multipole coefficients $a_\mathrm{E}(l,m)$ and $a_\mathrm{M}(l,m)$ characterize the weighting of the excited multipoles of the sphere, where the subscripts ``E” and ``M” distinguish between electric and magnetic multipoles. As for a known electric field $\mathbf{E}(\mathbf{r})$ around the scatterer, the scattering current density is $\mathbf{J}_{\rm s}(\mathbf{r})=-i\omega\left[\varepsilon(\mathbf{r})-\varepsilon_h\right]\mathbf{E}(\mathbf{r})$, where the $\varepsilon(\mathbf{r})$ is the permittivity at an arbitrary point $\mathbf{r}$, and $\varepsilon_h$ is the permittivity of the host medium. Using the Riccati-Bessel functions $\psi_l(kr)=krj_l(kr)$, the multipole coefficients can be expressed as \cite{grahnElectromagneticMultipoleTheory2012}:
\EqnTwo
\EqnThree
where $P_l^m$ are associated Legendre polynomials, $O_{lm}=\left[l(l+1)\right]^{-\frac{1}{2}} \left[\frac{(2l+1)}{4\pi}\frac{(l-m)!}{(l+m)!} \right]^{\frac{1}{2}}$, $\tau_{lm}(\theta)=d P_l^m(\cos \theta)/d\theta$ and $\pi_{lm} (\theta)=m P_l^m (\cos \theta)/\sin \theta $. We note that the multipole coefficients for a sphere can also be analytically obtained using Mie coefficients \cite{bohrenAbsorptionScatteringLight1998}. We implement the above numerical method in COMSOL for consistence with the visualization of C lines (to be discussed later). Using the above multipole coefficients, the total scattering cross section can be expressed as \cite{bohrenAbsorptionScatteringLight1998,grahnElectromagneticMultipoleTheory2012}:
\EqnFour
Equation\ (\ref{eqn:4}) can be employed to determine the contribution of individual multipole to the total scattering cross section. The contributions of the electric dipole, the magnetic dipole and the electric quadrupole are shown in Fig. \ref{fig:1}(a) as the dot-dashed red line, the dashed purple line, and the dashed yellow line. We will focus on the C lines of the electric field at two wavelengths $\lambda=950$ nm and $\lambda=1680$ nm, which are referred to as Type-I C line and Type-II C line, respectively. For the Type-I C line, the electric dipole, the magnetic dipole, and the electric quadrupole all contribute significantly to total scattering cross section, while for the Type-II C line the scattering cross section is dominated by the electric dipole moment. \textcolor{black}{Since the spatial structure of the C lines depends on the multipoles induced in the gold sphere, the proposed chiral discrimination method is frequency dependent, and there can be other types of C lines at different frequencies beyond the two types considered in this paper.}

The C lines can be directly visualized in COMSOL by determining the phase singularities of the scalar filed $\Psi=\mathbf{E}\cdot\mathbf{E}$ \cite{pengPolarizationSingularitiesLight2021}, and the results are shown in Fig. \ref{fig:1}(b) and (c) for Type-I and Type-II C lines, respectively. We notice that the C lines have similar structures in these two cases, i.e., a pair of C lines emerges in the near field and extends to the far-field region. To characterize the properties of the C lines for chiral discrimination, we then calculate the optical chirality of the C lines as \cite{tangOpticalChiralityIts2010,PhysRevLett.121.043901}:
\EqnFive
where $\mathcal{E}=\text{Re}(\mathbf{E})$ and $\mathcal{B}=\text{Re}(\mathbf{B})$. The ratio of ${\chi}$ and the electric field energy density $U_{\mathrm{e}}=\frac{1}{4} \varepsilon_{0}|\mathbf{E}|^{2}$  determines the degree of chiral asymmetry in the rate of excitation of a small chiral particle. The color of C lines in Fig. \ref{fig:1}(b) and (c) shows the value of $\chi/U_\text{e}$. We notice that ${\chi/U_\text{e}}$ is larger in the near field and its sign varies on the same C line. In addition, the value of ${\chi/U_\text{e}}$ is larger for the Type-I C line that has significant contribution from higher-order multipoles. \textcolor{black}{The emergence of optical chirality in this achiral system can be understood from the perspective of symmetry. The light propagation direction, the polarization vector, and the perpendicular displacement vector of the chiral particle form an orthogonal triad that can be right- or left-handed. Broadly, optical chirality can appear at spatial locations that are not on the symmetry planes (i.e., $xoz$- and $yoz$-planes, assuming the sphere locates at the origin).}

\FigTwo
\FigThree

\section{\label{sec: III. MST,SourceRep}	Absorption of a chiral particle at the Type-I C line }
We first consider the absorption of a chiral particle locating on the Type-I C line, as shown in Fig. \ref{fig:2}(a). Due to the symmetry of the C lines, we only consider the C line on the right-hand side of the sphere. The amplitude of the incident electric field is 1 V/m. The chiral particle is a two-pitch gold helix with minor radius $r = 5$ nm, major radius $R = 10$ nm, and pitch $p = 15$ nm, as shown in the inset in Fig. \ref{fig:2}(a). \textcolor{black}{We note that the chiral particle must be much smaller than the gold sphere so that the C lines will not be strongly perturbed by the scattering field of the chiral particle.} We study the absorption of the chiral particle while it moves along the C line, and totally 11 positions as marked by the black dots in Fig. \ref{fig:2}(a) are considered. The center axis of the helix is maintained  in the \textit{x} direction (i.e., the polarization direction of the incident plane wave).

We numerically calculated the absorption of the left-handed (LH) and right-handed (RH) helices. The results are denoted as the solid-blue line and dashed-red line in Fig. \ref{fig:2}(b), respectively. The absorption of LH helix reaches the maximum and minimum at the positions 4 and 8, respectively, while the absorption of the RH helix shows an opposite trend. At the center position 6 where optical chirality vanishes, the LH and RH helices give rise to equal absorption as expected. To characterize the different response of the LH and RH helices to the C line, we calculate the absorption dissymmetry factor as \cite{tangOpticalChiralityIts2010}:
\EqnSix
where $A^{\mathrm{L}}$ ($A^{\mathrm{R}}$) denotes the absorption of the LH (RH) helix. The numerical results of $g$ are shown in Fig.\ref{fig:2}(c) as blue circles. As seen, The dissymmetry factor $g$ reaches +1 and $-1$ at the positions 4 and 8 respectively, which are much larger than that under the excitation of a circularly polarized plane wave at the same wavelength and propagating in $z$ direction ($g = 0.28$ for the considered helix particle). \textcolor{black}{Figure \ref{fig:2}(d) shows the dissymmetry factor $g$ as a function of the distance from the gold sphere. As seen, as the distance increases, the absolute value of $g$ decreases except for the points near the symmetry plane where optical chirality vanishes (i.e., positions 5, 6, and 7). This is expected since the total field will gradually approach the linearly polarized incident field as the distance increases.}

To obtain intuitive understanding of the above numerical results, we apply multipole expansions to the helix particle and treat it approximately as a superposition of electric and magnetic dipoles due to the deep subwavelength of the particle size. Then, the absorption of the particle can be evaluated as:
\EqnSeven
where the Cartesian components of electric dipole moment \textbf{p} and magnetic dipole moment \textbf{m} are evaluated using $p_{i}=-\frac{1}{i \omega} \int_{V} J_{i} d v$ and $m_{i}=\frac{1}{2} \int_{V}(\mathbf{r} \times \mathbf{J})_{i} d v$ \cite{zhangOpticalForceToroidal2015}. Figure \ref{fig:3}(a) and (b) shows the amplitudes of the electric and magnetic dipoles induced in the LH (blue line) and RH (red line) helices at the considered positions, respectively. We notice that the dipoles of LH and RH helices are mirror-symmetric with respect to the center position 6. In addition, for either helix particle, both $|\mathbf{p}|$ and $|\mathbf{m}|$ are asymmetric with respect to the center position 6 due to the chirality of the particle. Interestingly, at the positions 4 and 8, $|\mathbf{p}|$ reaches local maximums, while $|\mathbf{m}|$ reaches local minimums. We apply Eqs. (\ref{eqn:6}) and (\ref{eqn:7}) to analytically evaluate the absorption dissymmetry factor, and the result is shown in Fig. \ref{fig:3}(c) as the solid line, which agrees well with the full-wave numerical result (symbols). This indicates that the absorption dissymmetry can be attributed to the difference of electric and magnetic dipoles induced in the chiral particles. 

\textcolor{black}{To further understand the underlying physical mechanism of the absorption dissymmetry, we decompose the dipoles as follows: 
\EqnEight
where $\overleftrightarrow{\alpha}_{\mathrm{e}}$, $\overleftrightarrow{\alpha}_{\mathrm{m}}$ and $\overleftrightarrow{\kappa}$ are electric polarizability, magnetic polarizability, and magnetoelectric polarizability of the helix, respectively, which are tensors depending on the orientation of the helix due to its anisotropic property. Using Eq.(\ref{eqn:7}), we can express the absorption difference of opposite helices as:
\EqnNine
Here, the contribution of the first term on the right-hand side is attributed to the off-diagonal elements of $\overleftrightarrow{\alpha}_{\mathrm{e}}^\mathrm{L,R}$ deriving from the anisotropic electric response of the helix. The second term is negligible since the helix is non-magnetic and deep subwavelength. The contribution of the third term is mainly attributed to the diagonal elements of $\overleftrightarrow{\kappa}^{\mathrm{L,R}}$ as $\frac{\omega}{2} \operatorname{Im}\left(\mathrm{i} 2 \mathbf{E}^{*} \cdot\left(\overleftrightarrow{\kappa}^{\mathrm{L}}-\overleftrightarrow{\kappa}^{\mathrm{R}}\right) \cdot \mathbf{B}\right) = -\sum_{i}\frac{4}{\varepsilon_{0}} \kappa_{i i}^{\prime \prime} \chi_{i}$, where $\chi_{i}=-\frac{\varepsilon_{0} \omega}{2} \operatorname{Im}\left(E_{i}^{*} B_{i}\right), i=x,y,z$ denotes the optical chirality contributed by each Cartesian component of the fields, and $\kappa_{ii}^{\prime \prime}$ denotes the imaginary part of the diagonal elements of $\overleftrightarrow{\kappa}^{\mathrm{L}}$. We numerically retrieved  $\overleftrightarrow{\alpha}_{\mathrm{e}}^\mathrm{L,R}$ and $\overleftrightarrow{\kappa}^{\mathrm{L,R}}$ based on the scattered far field of the helix, and then analytically evaluated Eq. (\ref{eqn:9}). The comparison of the absorption difference between the numerical and analytical results are shown as blue circles and red line in Fig. \ref{fig:4}(a), showing a good agreement. We notice that the contribution of the first term in Eq. (\ref{eqn:9}) (dashed yellow line) is much larger than that of the third term (dashed purple line). This is because that the electric polarizability is much larger than the magnetoelectric polarizability for the considered helix which is deep subwavelength. 
The denominator in Eq.(\ref{eqn:6}) can be expressed as 
\EqnTen
where again the magnetic contribution is neglected. We find that $\operatorname{Im}\left(\overleftrightarrow{\alpha}_{\mathrm{e}}^{\mathrm{L}}+\overleftrightarrow{\alpha}_{\mathrm{e}}^{\mathrm{R}}\right)$ is approximately isotropic. This is because that the off-diagonal elements of $\overleftrightarrow{\alpha}_{\mathrm{e}}^{\mathrm{L}}$ and $\overleftrightarrow{\alpha}_{\mathrm{e}}^{\mathrm{R}}$ cancel, while the diagonal elements are approximately equal since the helix has comparable length in three orthogonal directions. Thus, we can write $\operatorname{Im}\left(\overleftrightarrow{\alpha}_{\mathrm{e}}^{\mathrm{L}}+\overleftrightarrow{\alpha}_{\mathrm{e}}^{\mathrm{R}}\right) = 2 \alpha_{\mathrm{e}}^{\prime \prime} \overleftrightarrow{I}$, where $\alpha_{\mathrm{e}}^{\prime \prime}$ denotes the imaginary part of the diagonal elements and $\overleftrightarrow{I}$ is the unit tensor. Equation (\ref{eqn:10}) is reduced to
\EqnEleven
where $U_{\mathrm{e}}=\frac{1}{4} \varepsilon_{0}|\mathbf{E}|^{2}$ is the time-averaged electric field energy density. Using Eqs. (\ref{eqn:9}) and (\ref{eqn:11}), we can rewrite Eq. (\ref{eqn:6}) as
\EqnTwelve
The above equation shows that the dissymmetric factor $g$ is determined by both the electric polarizability and the magnetoelectric polarizability, which are anisotropic. Thus, $g$ must depend on the orientation of the helix and is generally not proportional to the normalized total optical chirality $\chi / U_{\mathrm{e}}=\sum_{i} \chi_{i} / U_{\mathrm{e}}$. We note that for each Cartesian component of the second term, it is still proportional to the corresponding component of normalized optical chirality (i.e., $\chi_{i} / U_{\mathrm{e}}$). For an isotropic chiral particle, Eq. (\ref{eqn:12}) reduces to $g =-\frac{2 \kappa^{\prime \prime} \chi}{\omega \alpha_{\mathrm{e}}^{\prime \prime} U_{\mathrm{e}}}$ (i.e., the expression in Ref. \cite{tangOpticalChiralityIts2010}).}

\textcolor{black}{To understand the dependence of $g$ on the optical chirality, we plot in Fig. \ref{fig:4}(b) the dissymmetry factor contributed by the second term in Eq. (\ref{eqn:12}) for each Cartesian component $i=x,y,z$, as denoted by dashed red lines. The sold blue lines denote the corresponding component of the normalized optical chirality. As seen, each component of the dissymmetry factor (i.e., $g_{\kappa i}$) agrees well with the normalized optical chirality $\chi_{i} / U_{\mathrm{e}}$. In addition, both the dissymmetry factor and the normalized optical chirality are different for the three components, demonstrating the anisotropy of the helix and the field. This indicates that the absorption dissymmetry is sensitive to the orientation of the helix, which may be employed for experimentally characterizing the orientation of anisotropic chiral particles/molecules.} 

 \FigFour

\section{\label{sec: IV. MST,SourceRep}		Absorption of a chiral particle at the Type-II C line }
The Type-I C line is contributed by the electric dipole, magnetic dipole and electric quadrupole induced in the gold sphere. To understand how the different multipole components of a C line affects the optical chirality and the absorption of a chiral particle, we then consider the configuration in Fig. \ref{fig:5}(a), where the helix particle locates at the Type-II C line that is dominated by electric dipole moment. Similarly, we calculated the absorption of the same LH and RH helices located as different positions of the C line, and the results are shown in Fig. \ref{fig:5}(b). In contrast to the case of Type-I C line, the absorption of the helices are more symmetric in this case. The absorption of the LH helix reaches the minimum value at position 7, while the minimum absorption of the RH helix is at position 5.

\FigFive
\FigSix

 \textcolor{black}{We calculated the absorption dissymmetry for this case using Eq. (\ref{eqn:6}), and the results are shown in Fig. \ref{fig:5}(c). We notice that the dissymmetry factor $g$ of the Type-II C line is only slightly larger than that under the excitation of a circularly polarized plane wave propagating in $z$ direction and is smaller than that of the Type-I C line in Fig. \ref{fig:2}(c). This distinguishes the two types of C lines in terms of the ability of chiral discrimination. Figure \ref{fig:5}(d) shows the dissymmetry factor $g$ for the considered 11 positions as a function of distance from the sphere. Similar to Fig. \ref{fig:2}(d), its absolute value decreases as the distance is increased except for the position 6 locating on the symmetry plane where optical chirality vanishes.}
 
We analytically calculated the amplitudes of the electric and magnetic dipole moments induced in the helices at the considered positions of the Type-II C line. The results are shown in Fig. \ref{fig:6}(a) and (b), where the blue lines denote the results of the LH helix and red dashed line denotes the results of the RH helix. Similar to the previous case, the dipoles of the LH and RH helices are symmetric with respect to the center position 6, while the dipoles of individual helix are asymmetric due to the chirality of the particle. In addition, the electric and magnetic dipoles show similar trend. Using Eqs. (\ref{eqn:6}) and (\ref{eqn:7}), we analytically calculate the absorption dissymmetry factor. The result is shown in Fig. \ref{fig:6}(c) as the solid line. We notice that the analytical result well agrees with the full-wave numerical result (symbols). This again demonstrates the validity of the multipole expansions method.

\FigSeven

\textcolor{black}{Using Eq. (\ref{eqn:9}), we then analytically evaluate the contributions to the absorption difference by different polariabilities of the helix. The results are shown in Fig. \ref{fig:7}(a), where the red line denotes the analytical result of the total absorption difference, which agrees well with the numerical result denoted by the blue circles. The dashed yellow line denotes the absorption difference contributed by the electric polarizability, while the dashed purple line denotes the contribution of the magnetoelectric polarizability. Similar to the case of Type-I C line, the absorption difference is mainly contributed by the electric polarizability due to the weak magnetoelectric effect of the helix at deep subwavelength.}

\textcolor{black}{To understand the dependence of the absorption dissymmetry factor $g$ on the optical chirality, we apply Eq. (\ref{eqn:12}) to evaluate the dissymmetry contributed by each Cartesian component of the optical chirality. The results are shown in Fig. \ref{fig:7}(b) as the dashed red lines. The sold blue lines denote the corresponding component of the normalized optical chirality. Similar to the case of the Type-I C line, the absorption dissymmetry are different for the three components, but they all agree well with the normalized optical chirality $\chi_{i} / U_{\mathrm{e}}$.}

\section{\label{sec: V.	Conclusion} Discussion and Conclusion}
In summary, we propose a new method to detect chiral particles by using the C lines of a gold sphere excited by linearly polarized light.  We show that the C lines can give rise to larger dissymmetry factor of absorption compared to normal plane waves of circular polarization. Importantly, we find that the multipole origin of the C lines can affect the dissymmetry factor and thus the ability of chiral discrimination. The C lines with higher-order multipoles can give rise to larger dissymmetry factor compared to the C lines dominated by electric dipole. By treating the helix particles as electric and magnetic dipoles, we uncover the interaction of the chiral particles with the C lines and its contribution to the absorption dissymmetry. The  electric and magnetic dipoles have different amplitudes for the LH and RH helices, which account for their different absorption properties. \textcolor{black}{The absorption dissymmetry is attributed to both the anisotropic electric polarizability and the anisotropic magnetoelectric polarizability of the helix and thus is sensitive to the orientation of the helix.} 

\textcolor{black}{To implement the proposed approach of chiral discrimination, one can measure the absorption rate by comparing the intensity of light with and without the chiral particle \cite{celebrano2011single,vinegradCircularDichroismSingle2018}, and the absorption dissymmetry can then be determined by placing opposite chiral particles at the same positions on the C line.
A limitation of the proposed method is that it depends on the position of the chiral particle and cannot apply to a homogenous medium such as a chiral fluid. This can be overcome by employing periodic structures such as metasurfaces to induce volumetric optical chirality. The specific design and the involved physics require further studies.} 

Our study provides a new method for chiral discrimination without using chiral excitations or chiral structures. Since a metal sphere is one of the simplest structures in nano fabrications, the proposed method should be easier to implement compared to previous proposals that involve complex structures and delicate designs. Our results can generate novel applications in optical sensing, optical manipulations, and chiral quantum optics. 

\section{\label{sec: VII. Acknowledgements}Acknowledgements}
The work described in this paper was supported by grants from the Research Grants Council of the Hong Kong Special Administrative Region, China (Project No. CityU 11306019) and the National Natural Science Foundation of China (Project No. 11904306) . 
\bibliography{references_mendeley}
\end{document}